\documentclass[%
 reprint,
 amsmath,amssymb,
 aps,
]{revtex4-1}
\usepackage{verbatim}
\usepackage{graphicx}
\usepackage{dcolumn}
\usepackage{bm,amsmath}

\usepackage{verbatim}
\usepackage{graphicx}
\usepackage{dcolumn}
\usepackage{bm}

\usepackage{float}                  
\usepackage{subfig}                 
\usepackage{overpic}                

\begin{document}

\preprint{APS/123-QED}

\title{A lattice Boltzmann scheme  with  arbitrary Prandtl number and \\ specific heat ratio based on the polyatomic ES-BGK model}

\author{Kainan Hu}
\email{Corresponding author: hu$\_$kainan@outlook.com}
\altaffiliation{East China University of Technology, Nanchang, China}

\author{Hongwu Zhang}
\affiliation{
Industrial Gas Turbine Laboratory, Institute of Engineering Thermophysics,\\
 Chinese Academy of Sciences, Beijing, China
}

\author{Juan Du}
\affiliation{
Industrial Gas Turbine Laboratory, Institute of Engineering Thermophysics,\\
 Chinese Academy of Sciences, Beijing, China
}

\author{Yan Jin}
\affiliation{
Center of Applied Space Technology and Microgravity, University of Bremen, \\
 Bremen, Germany
}
\author{Hui Kong}
\affiliation{
School of Mechanical Engineering, Beijing Institute of Technology, Beijing, China
}

\date{\today}


\begin{abstract}

A Boltzmann lattice scheme based on the polyatomic ellipsoidal statistical model (ES-BGK) is proposed, incorporating an arbitrary  Prandtl number and a specific heat ratio. The Prandtl number is modulated by a parameter in the Gaussian distribution, while the specific heat ratio is adjusted by additional free degrees. The Gaussian distribution is expanded using Hermite polynomials, and a general formula for the Hermite coefficients of the Gaussian distribution is derived. Benchmark tests are conducted to validate the proposed scheme, with numerical results demonstrating good agreement with analytical solutions.

\begin{description}
\item[PACS numbers]
47.11.-j, 47.10.-g, 47.40.-x
\end{description}
\end{abstract}

\maketitle


\section{\label{sec:sec1}Introduction}

Over the past three decades, the Lattice Boltzmann Method (LBM) has been widely applied to a variety of complex flows, such as multiphase and multicomponent fluids \cite{guo2005finite}\cite{li2016lattice}, fluid flows through porous media \cite{ferreol1995lattice}, turbulence\cite{chen2003extended}, magneto-hydrodynamics\cite{rahmati2016double}, and thermal fluids\cite{yoshino2004comparison}. However, when LBM is applied to thermal fluids, it encounters some issues. Specifically, the Navier-Stokes equations (including the energy conversation equation) can be derived from the evolution equation via the Chapman-Enskog expansion, but the specific heat ratio and the Prandtl number are fixed. As a result, the transport coefficients obtained this way are not realistic for polyatomic molecule fluids.

To address this issue, some LB schemes combined with arbitrary  specific heat ratio or Prandtl number have been proposed. To modify the specific heat ratio, new variables such as rotational velocity or rotational energy are introduced and discretized in the velocity space \cite{kataoka2004lattice,watari2007finite,tsutahara2008new,nie2008thermal}. While these LB schemes have clear physical meanings and are easy to implement, they require the setting of parameters through experience. To modify the Prandtl number, a quasi-equilibrium or other similar element that controls the Prandtl number is introduced \cite{frapolli2014multispeed,chen1997two,yan2011flux,soe1998thermal}. Some works give arbitrary specific heat ratio and Prandtl number at the same time (Dellar, 2008; Saadat et al., 2019).

Most of these schemes are based on the BGK collision model, which assumes that the molecule is monatomic.
An alternative collision model, the Ellipsoidal Statistics Model (ES-BGK) \cite{holway1965kinetic}, has been proposed. The model has additional parameters, yielding satisfactory transport coefficients at the Navier-Stokes level and satisfying the entropy inequality ($H$-theorem) \cite{andries2000gaussian,andries2001bgk,andries2002numerical,brull2009ellipsoidal,meng2013lattice}, which is an essential requirement for a successful LB scheme.

In this work, we present a novel LB scheme with an arbitrary  Prandtl number (and an arbitrary  specific heat ratio) based on the ES-BGK model. 
Our main contribution was deriving a general term formula for the Hermite coefficients of the Gaussian distribution, which facilitates the derivation of high-order terms of the Hermite coefficients. This general term formula makes the design of high-order LB schemes based on the ES-BGK model more straightforward and efficient. Consequently, the general term formula can be used to efficiently design LB schemes applied to 2D or 3D flows.
Compared to existing schemes, the proposed LB scheme has several advantages. Its physical meaning is clear and it adheres to the $H$-theorem, which is essential for a lattice Boltzmann scheme. Additionally, it does not require parameters to be set through experience, and its implementation is much simpler, as it only requires rewriting the code of the equilibrium distribution function, while the other parts of the code and the basic structure remain unchanged.

\section{Polyatomic ES-BGK  model}

This section discusses the polyatomic ES-BGK model, which is characterized by the kinetic equation of the polyatomic distribution function $f(\bm{\xi},\bm{x},t, I)$\cite{andries2000gaussian, brull2009ellipsoidal}.
\begin{equation}\label{Eq:Boltzmann}
\frac{\partial f}{\partial t} + \bm{\xi}\cdot\frac{\partial f}{\partial \bm{x}} = \Omega,
\end{equation}
where $f(\bm{\xi},\bm{x},t,I)$ is the distribution function with the particle velocity $\bm{\xi}$ and the internal energy $\epsilon(I) = I^{\frac{2}{N_f}}$ at position $\bm{x}$ and time $t$, $N_f$ being the additional degrees of freedom of the gas. The collision operator $\Omega(f)$ is given by
\begin{equation}
\Omega(f) = -\frac{1}{\tau(1-\nu+\nu\kappa)}(f-G[f]),
\end{equation}
where $\tau$ is the relaxation time and $G[f]$ is the polyatomic Gaussian model. The parameters $-\frac{2}{3}\leqslant \nu<1$ and $0\leqslant \kappa\leqslant 1$ are introduced to modify the Prandtl number and the specific heat ratio.

The density $\rho$, macroscopic velocity $\bm{u}$, and total energy $E$, as well as the specific internal energy $e$, are defined by the moments of the distribution function $f$ as follows:
\begin{subequations}
\begin{align}
\rho = & \int_{\mathbb{R}^D}\!\! \int_{\mathbb{R}^+} f \, d\bm{\xi} \, dI,\\
\rho\bm{u} =& \int_{\mathbb{R}^D}\!\!\int_{\mathbb{R}^+}  \bm{\xi} f \, d\bm{\xi} \, dI,\\
E = \frac{1}{2}\rho u^2 + \rho e  =& \int_{\mathbb{R}^D}\!\!\int_{\mathbb{R}^+}(\frac{1}{2}\xi^2 + I^{\frac{2}{N_f}}) f \, d\bm{\xi} \, dI,
\end{align}
\end{subequations}
where $D$ is the space dimension.

The specific internal energy $e$ can be expressed as the sum of two components: the internal energy of translational velocity $e_{tr}$ and the energy associated with the internal structure $e_{int}$, as follows:
\begin{equation}
e = e_{tr} + e_{int},
\end{equation}
where
\begin{align}
e_{tr} &= \frac{1}{\rho} \int_{R^D} \!\!\int_{R^+} \frac{1}{2}(\bm{\xi} - \bm{u})^2 f d\bm{\xi} d{I},\\
e_{int} &=\frac{1}{\rho} \int_{R^D} \!\!\int_{R^+} I^{2/N_f} f d\bm{\xi} d{I}.
\end{align}
The relationship between temperature ($T$, $T_{tr}$, $T_{int}$) and the corresponding energy ($e$, $e_{tr}$, $e_{int}$) can be expressed as follows:
\begin{align}
e = \frac{D+N_f}{2}R_g T_{eq},\quad e_{tr} = \frac{D}{2}R_g T_{tr}, \quad e_{int} = \frac{N_f}{2} R_g T_{int},
\end{align}
where $R_g$ is the universal gas constant.

The state equation can be expressed as $p=\rho R_g T_{eq}$, where $p$ denotes the pressure.

The Generalized Gaussian Model is defined by the equation:

\begin{align}
 G[f] = & \frac{\Lambda_{N_f} \rho }{\sqrt{det(2\pi \bold{\Lambda)}}}\frac{1}{ R_g T_{rel}^{N_f/2}}\notag\\ 
 &\times\exp{\Big[-\frac{1}{2}(\bm{\xi} - \bm{u})\cdot \bm{\Lambda}^{-1} \cdot (\bm{\xi} -\bm{u})
  - \frac{I^{\frac{2}{N_f}}}{R_g T_{rel}} \Big]},
\end{align}

where $\bm{\Lambda}$ is the corrected tensor given by
\begin{equation}
\bm{\Lambda}= (1-\kappa)[(1-\nu)R_g T_{tr} \bm{\delta} + \nu \bm{\sigma}] + \kappa R_g T_{eq} \bm{\delta},
\end{equation}
$\bm{\sigma}$ is the opposite stress tensor,
\begin{equation}\label{Eq:Lambda}
\bm{\sigma} = \frac{1}{\rho}\int (\bm{\xi}-\bm{u}) (\bm{\xi}-\bm{u}) f d \bm{\xi},
\end{equation}
and $\bm{\delta}$ is the unit tensor.
The relaxation temperature $T_{rel}$ is defined as
\begin{equation}
T_{rel} = \kappa T_{eq} + (1-\kappa)T_{int},
\end{equation}
and the constant $\Lambda_{N_f}$ is given by
\begin{equation}
\Lambda_{N_f}^{-1}\!=\!\int e^{-I^2/N_f}d I.
\end{equation}

\section{Polyatomic ES-BGK model: Description with two distribution functions}
The evolution equation of the polyatomic distribution function $f(\bm{\xi},\bm{x},t,I)$ can be expressed as
\begin{equation}\label{Eq:EvolF}
\frac{\partial}{\partial t} f +\bm{\xi}\cdot \frac{\partial}{\partial \bm{x}} f
                            = -\frac{1}{\tau(1-\nu+\kappa \nu)} \left( f - G[f] \right),
\end{equation}
where $\tau$, $\nu$, and $\kappa$ are constants.
The kinetic equation can be reduced to two distribution functions, namely the mass distribution function $g(\bm{\xi})$ and the energy distribution function $h(\bm{\xi})$, as proposed by C.K.Chu \cite{chu1965kinetic,chu1965kinetic2} and V.A.Rykov \cite{rykov1975model}. This approach has the advantage of reducing computational resources, as well as eliminating the need to discretize $\epsilon$ in the discrete velocity space when applied to the Lattice Boltzmann Method (LBM). The lattices employed in this scheme are DnQb models, which are simpler to design than those requiring the discretization of both the translational velocity of the particle and the newly introduced parameter \cite{watari2007finite,tsutahara2008new,yudistiawan2010higher,yan2011flux}. 

The distributions of $g(\bm{\xi},\bm{x},t)$ and $h(\bm{\xi},\bm{x},t)$ are defined by
\begin{subequations}
\begin{align}
g(\bm{\xi},\bm{x},t) = &\int_{R^{+}} f(\bm{\xi},\bm{x},t,I) \mathrm{d} I, \\
h(\bm{\xi},\bm{x},t) = &\int_{R^{+}} I^{\frac{2}{N_f}} f(\bm{\xi},\bm{x},t,I) \mathrm{d} I.
\end{align}
\end{subequations}
The macroscopic quantities are determined by the moments of the mass distribution $g(\bm{\xi})$ and the energy distribution $h(\bm{\xi})$ as follows:
\begin{subequations}
\begin{align}
\rho = &\int_{R^{D}} g(\bm{\xi}) \mathrm{d} \bm{\xi},\\
\rho \bm{u} =& \int_{R^{D}} \bm{\xi} g(\bm{\xi}) \mathrm{d} \bm{\xi},\\
\rho E = &\int_{R^{D}} [ \frac{1}{2}\xi^2 g(\bm{\xi}) + h(\bm{\xi}) ] \mathrm{d} \bm{\xi}.
\end{align}
\end{subequations}
Integrating Eq.(~~\ref{Eq:EvolF}) on $I$, we obtain the evolution equation of $g(\bm{\xi})$
\begin{align}\label{Eq:ReductionG}
\frac{\partial }{\partial t}g + \bm{\xi}\cdot\nabla g = -\frac{1}{\tau(1-\nu +\kappa \nu)}(g -  G[g,h]),
\end{align}
where $G[g,h]$ is
\begin{align}\label{Eq:Gf}
G[g,h]=\frac{\rho}{\sqrt{\mathrm{det}(2\pi \bold{\Lambda})}}\exp \left[-\frac{1}{2}(\bm{\xi} - \bm{u})\cdot \bm{\Lambda}^{-1} \cdot (\bm{\xi} -\bm{u})\right].
\end{align}

Integrating Eq.~(~\ref{Eq:EvolF}) multiplied by $I^{\frac{2}{N_f}}$ over $I$, we obtain the evolution equation of $h(\bm{\xi})$ as

\begin{align}\label{Eq:ReductionH}
\frac{\partial }{\partial t}h + \bm{\xi}\cdot\nabla h = -\frac{1}{\tau(1-\nu +\kappa \nu)}(h - \frac{N_f}{2}
R_g T_{\mathrm{rel}} G[g,h]).
\end{align}

From Eq.(~\ref{Eq:ReductionG}) and (~\ref{Eq:ReductionH}), the Navier-Stokes equations with an arbitrary  specific heat ratio and Prandtl number can be derived via the Chapman-Enskog expansion \cite{andries2000gaussian,ansumali2003minimal}. In the derived Navier-Stokes equations, the viscosity tensor is given by $\bm{\sigma} = \mu(\nabla \bm{u} + \bm{u} \nabla - \alpha \nabla \cdot \bm{u} \bm{\delta})$, where $\mu = \tau p/(1-\nu + \kappa \nu)$ is the viscosity coefficient and $\mu\alpha$ is the second viscosity coefficient, with $\alpha = (\gamma-1)-\frac{1-\kappa}{\kappa}(1-\nu)(\frac{D+2}{D}-\gamma)$.
The specific heat ratio $\gamma$ and Prandtl number $Pr$ in the recovered Navier-Stokes equations are defined, respectively, as:
\begin{align}
\gamma = \frac{N_f + D +2}{N_f +D}\label{Eq:Gamma},\\
Pr = \frac{1}{1-\nu + \kappa \nu}\label{Eq:Pr},
\end{align}
where $2/3 \leq Pr < +\infty$.

\section{The General Term Formula for the Hermite Coefficients  of  the Gaussian distribution}\label{HermiteG}
The expansion of the Maxwell-Boltzmann distribution on the Hermite polynomials has been explored in the literature \cite{grad1949note,shim2013obtain,mattila2014high}. Here, we extend this discussion, expand the Gaussian distribution on the Hermite polynomials.

The identities given by Grad in \cite{grad1949note}, i.e. Eq. (4), (12) and (16), are employed in the following paragraphs. These are:
\begin{align}
(\bm{x}+\bm{y})^n &= \sum^{n}_{r=0}\bm{x}^r \bm{y}^{n-r}, \\\mu_{2n} &= \int \omega(\bm{x})\bm{x}^{2n}d\bm{x}=\bm{\delta}^n ,\\\bm{H}^{(n)}(\bm x) &= \bm{x}^n - \bm{\delta}\bm{x}^{n-2} + \bm{\delta}^2\bm{x}^{n-4}+\cdots,   
\end{align}
where $\bm{H}^{(n)}(\bm x)$ is the Hermite polynomials and $\omega(\bm x) $ is the weight function
$$\omega(\bm x) = \displaystyle\frac{1}{\sqrt{(2\pi)^{D/2}}}\exp{(-\frac{x^2}{2})}.$$
The following equations are derived from the above identities:
\begin{align}
\bm{H}^{(0)}(\bm \xi) &= 1,\notag\\
\bm{H}^{(1)}(\bm \xi) &= \bm \xi , \notag\\
\bm{H}^{(2)}(\bm \xi) &= \bm \xi^2 - \bm{\delta}, \notag\\
\bm{H}^{(3)}(\bm \xi) &= \bm \xi^3 -  \bm{\delta}\bm{\xi},\notag\\
\bm{H}^{(4)}(\bm \xi) &= \bm \xi^4 -  \bm{\delta}\bm{\xi}^2 + \bm{\delta}^2,\notag\\
\bm{H}^{(5)}(\bm \xi) &= \bm \xi^5 -  \bm{\delta}\bm{\xi}^3 + \bm{\delta}^2\bm \xi, \notag \\
\bm{H}^{(6)}(\bm \xi) &= \bm \xi^6 -  \bm{\delta}\bm{\xi}^4 + \bm{\delta}^2\bm \xi ^2 -\bm{\delta}^3 \notag
\end{align}
The identities given by Grad in \cite{grad1949note}, i.e. Eq. (4), (12) and (16), are employed in the following equations. These equations are derived from the identities, which involve the Hermite polynomials $\bm{H}^{(n)}(\bm x)$ and the weight function $\omega(\bm x)$.

The Gaussian distribution $G$ can be expressed as an expansion on the Hermite polynomial, as shown in Equation Eq.(~\ref{Eq:fesHerm}):
\begin{equation}\label{Eq:fesHerm}
  G=\rho\omega(\bm{\xi})\sum_{n=0}^{\infty} \frac{1}{n!}\bm{a}^{(n)}\cdot \bm{H}^{(n)}(\bm{\xi}),
\end{equation}
where $\bm{a}^{(n)}$ are the expansion coefficients and $\bm{H}^{(n)}(\bm{\xi})$ are the Hermite polynomials.
The expansion coefficients $\bm{a}^{(n)}$ can be obtained by 
\begin{equation}\label{Eq:HermiteCo}
  \bm{a}^{(n)} = \frac{1}{\rho}\int \bm{H}^{(n)}(\bm{\xi})G d \bm{\xi}.
\end{equation}

By defining $(\bm{\xi} - \bm{u}) \cdot \bm{\Lambda}^{-\frac{1}{2}} =\bm{\eta}$, we obtain $\bm{\xi}=\bm{u} + \bm{\eta} \cdot \bm{\Lambda}^{\frac{1}{2}}$. Inserting this and Eq.(~\ref{Eq:Gf}) into Eq.(~\ref{Eq:HermiteCo}), we obtain
\begin{align}\label{Eq:HermiteCo2}
 \bm{a}^{(n)} = \sum_{\alpha=0}^{[n/2]} \sum_{\beta =0}^{[n/2]-\alpha}  (-1)^{\alpha}  \bm{\delta}^{\alpha} \bm{\Lambda}^{\beta} \bm{u}^{n-2(\alpha +\beta)}.
\end{align}
The derivation of Eq.~(\ref{Eq:HermiteCo2}) can be found in Appendix.~(\ref{appendix:A}).

Defining $\chi\!=\!\alpha + \beta $, inserting $\chi$ into Eq({~\ref{Eq:HermiteCo2}) and changing the ranges of the superscripts  we obtain
\begin{align}
 \bm{a}^{(n)} =  &\sum_{\chi=0}^{[n/2]}\sum_{\beta=0}^{\chi} (-1)^{\beta}\bm{\Lambda}^{\beta} \bm{\delta}^{\chi-\beta} \bm{u}^{n-2\chi}\notag\\
   =&\sum_{\chi=0}^{[n/2]}(\bm{\Lambda}-\bm{\delta)}^{\chi}\bm{u}^{n-2\chi}\notag\\
   =&\sum_{\alpha=0}^{[n/2]}(\bm{\Lambda}-\bm{\delta)}^{\alpha}\bm{u}^{n-2\alpha}.\notag
\end{align}


Finally, we obtain the general term formula for the Hermite coefficients of the Gaussian distribution $G$
\begin{align}\label{Eq:fESCoe}
 \bm{a}^{(n)}=\sum_{\alpha=0}^{[n/2]}\sum_{\beta=0}^{\alpha}(-1)^{\alpha - \beta}\bm{\Lambda}^{\beta}\bm{\delta}^{\alpha - \beta}\bm{u}^{n-2\alpha}.
\end{align}
The first six orders of $\bm{a}^{(n)}$ are provided
\begin{align}
  \bm{a}^{(0)} &=1 ,\notag \\
  \bm{a}^{(1)} &=\bm{u},\notag\\
  \bm{a}^{(2)} &=\bm{\Lambda}-\bm{\delta} +\bm{u}^2,\notag\\
  \bm{a}^{(3)} &=(\bm{\Lambda}-\bm{\delta})\bm{u} +\bm{u}^3,\notag\\
  \bm{a}^{(4)} &=(\bm{\Lambda}-\bm{\delta})^2 +(\bm{\Lambda}-\bm{\delta})\bm{u}^2+\bm{u}^4,\notag\\
  \bm{a}^{(5)} &=(\bm{\Lambda}-\bm{\delta})^2\bm{u} +(\bm{\Lambda}-\bm{\delta})\bm{u}^3+\bm{u}^5,\notag\\
\bm{a}^{(6)}&=(\bm{\Lambda}-\bm{\delta})^3+(\bm{\Lambda}-\bm{\delta})^2\bm{u}^2+(\bm{\Lambda}-\bm{\delta})\bm{u}^4+\bm{u}^6.\notag
\end{align}

If we wish to recover the Burnet equations through the Chapman-Enskog expansion, the sixth order of $\bm{a}^{(n)}$ is required. In this study, we focus on equilibrium flow, for which only the fourth order of the Hermite expansion of the Gaussian distribution is necessary, as expressed by the following equation,
\begin{equation}\label{Eq:GExpansion}
G^{(4)}=\rho\omega(\bm{\xi})\sum_{n=0}^{4} \frac{1}{n!}\bm{a}^{(n)}\cdot \bm{H}^{(n)}(\bm{\xi}).
\end{equation}
The expansion of Eq.~(\ref{Eq:GExpansion}) is in  Appdenix.~(\ref{appendix:B}).

\section{Lattice Boltzmann scheme based on the ES-BGK model}
For convenience, we introduced the dimensionless variables,

\begin{align}\label{Eq:DDF}
\tilde f^{eq}(\tilde{\bm{\xi}},\tilde{I},\tilde{\bm{x}},\tilde{t}) =& \tilde \Lambda_\delta \frac{\tilde \rho}{(2 \pi \tilde \theta)^\frac{D}{2}}
\frac{1}{\tilde \theta^{\frac{\delta}{2}}} \notag\\
&\times \exp  \Big(-\frac{|\tilde {\bm{\xi}}- \tilde {\bm{u}}|^2}{2 \tilde{\theta}}\Big) \exp \Big( -\frac{\tilde I^{\frac{2}{\delta}}}{\tilde \theta }\Big),
\end{align}

where
\[ \begin{array}{ccccc}
&\tilde{\bm{x}}=\displaystyle\frac{\bm{x}}{L_0},
&\tilde{\bm{u}}=\displaystyle\frac{\bm{u}}{\sqrt{\theta_0}},
&\tilde{\bm{\xi}}=\displaystyle\frac{\bm{\xi}}{\sqrt{\theta_0}},
&\tilde \rho=\displaystyle\frac{\rho}{ \rho_0},
\end{array}\]
the dimensionless formation of the equilibrium distribution function is obtained.

The following passage presents a set of dimensionless variables, where $L_0$, $T_0$, $\rho_0$, $t_0$, and $\theta_0 = R_g T_0$ are the characteristic length, temperature, density, time, and thermal energy, respectively. All variables are then expressed in terms of these characteristics:
\[\begin{array}{llll}
&\bm{x} = \displaystyle\frac{\bm{x}}{L_0},
&\bm{u} = \displaystyle\frac{\bm{u}}{\sqrt{\theta_0}},
&\bm{\xi} = \displaystyle\frac{\bm{\xi}}{\sqrt{\theta_0}},\\
&\rho = \displaystyle\frac{\rho}{ \rho_0},
&T_{eq} = \displaystyle\frac{T_{eq}}{T_0},
&T_{tr} = \displaystyle\frac{T_{tr}}{T_0},\\
&T_{rel} = \displaystyle\frac{T_{rel}}{T_0},
&T_{int} = \displaystyle\frac{T_{int}}{T_0},
&E = \displaystyle\frac{E}{\theta_0},\\
&e = \displaystyle\frac{e}{\theta_0},
&e_{tr} = \displaystyle\frac{e_{tr}}{\theta_0},
&e_{int} = \displaystyle\frac{e_{int}}{\theta_0},\\
&p = \displaystyle\frac{p}{(\rho_0\theta_0)},
&I = \displaystyle\frac{I}{\theta_0^{N_f/2}},
&G[g,h] = \displaystyle \frac{ G[g,h]\displaystyle\theta_0^{D/2}}{\rho_0},\\
&g = \displaystyle\frac{g\displaystyle\theta_0^{D/2}}{\rho_0},
&h = \displaystyle\frac{h\displaystyle\theta_0^{(D+N_f)/2}}{\rho_0},
&\Lambda_{\tiny{{N_f}}}^{-1} = \int \exp( -I^{\frac{2}{N_f}}) d I\\
&t = \displaystyle\frac{t\sqrt{(RT_0)}}{L_0},
&\tau = \displaystyle\frac{\tau\sqrt{(RT_0)}}{L_0},
&\mu = \displaystyle\frac{\mu}{\rho_0 L_0 \sqrt{(RT_0)}}.\\
\end{array}\]
In the following part, all the variables are dimensionless and the tildes are omitted.

Discretizing $\bm{\xi}$,$g$,$h$,$G[g,h]$ in the discrete velocity space, we get $\bm{\xi}_i$,$g_i$,$h_i$ and $G_i[g,h]$.
 The discrete distribution $g_i$, $h_i$ and the Gaussian distribution $G_i[g,h]$ are defined by
\begin{align}
g_i = \frac{\omega_i g(\bm{x},\bm{\xi}_i,t)}{\omega({\bm{\xi}_i})},
h_i = \frac{\omega_i h(\bm{x},\bm{\xi}_i,t)}{\omega({\bm{\xi}_i})},
G_i = \frac{\omega_i G(\bm{x},\bm{\xi}_i,t)}{\omega({\bm{\xi}_i})}.\notag
\end{align}

After discretizing the evolution equations of $g$ and $h$ (i.e., Eq. (~\ref{Eq:ReductionG}) and (~\ref{Eq:ReductionH})), and inserting the definition of Prandlt number, we obtain the discrete evolution equations of $g_i$ and $h_i$:
\begin{subequations}\label{Eq:DisReductionHA}
\begin{align}
\frac{\partial }{\partial t}g_i + \bm{\xi}_i\cdot\nabla g_i &= -\frac{Pr}{\tau}(g_i - G_{i}[g,h]),\label{Eq:DisReductionG}\\
\frac{\partial }{\partial t}h_i + \bm{\xi}_i\cdot\nabla h_i &= -\frac{Pr}{\tau}(h_i - \frac{N_f}{2} T_{\text{eq}} G_{i}[g,h]).\label{Eq:DisReductionH}
\end{align}
\end{subequations}

In discrete velocity space, the density $\rho$, the macroscopic velocity $\bm{u}$, and the specific total energy $E$ are defined by the following equations:
\begin{subequations}
\begin{align}
\rho =& \sum_i g_i, \\
\rho \bm{u} =& \sum_i g_i \bm{\xi}_i, \\
\rho E =& \sum_i g_i \frac{1}{2}\xi^2_i + \sum_i h_i.\label{Eq:TransEnergy}
\end{align}
\end{subequations}

Moreover, the following relationships hold:
\begin{subequations}
\begin{align}
\rho \left(\frac{1}{2}u^2 + e_{tr}\right) =& \sum_i g_i \frac{1}{2}\xi^2_i, \\
\frac{1}{2}\rho e_{int} =& \sum_i h_i.
\end{align}
\end{subequations}

The dimensionless state equation is $p=\rho T$, and the relationships between the dimensionless temperatures $T_{eq}$, $T_{tr}$, $T_{int}$ and the corresponding dimensionless energies $e$, $e_{tr}$, $e_{int}$ are
\begin{align}
e = \frac{D+N_f}{2} T_{eq},\quad e_{tr} = \frac{D}{2} T_{tr}, \quad e_{int} = \frac{N_f}{2} T_{int}.
\end{align}
The dimensionless correct tensor is
\begin{equation}
\bm{\Lambda}= (1-\kappa)[(1-\nu)T_{tr} \bm{\delta} + \nu \bm{\sigma}] + \kappa  T_{eq} \bm{\delta}.
\end{equation}
The reduced evolution of $g$, as expressed in Eq.~(\ref{Eq:DisReductionG}), is discretized along the characteristics direction, 
yielding the following equation
\begin{align}\label{Eq:DiscreteReductionG}
&g_i(\bm{x} + \bm{\xi}_i \Delta t,t + \Delta t)\notag\\
 &= g_i(\bm{x},t)  -  \frac{Pr}{\tau_f}[g_i(\bm{x},t) -G_i(\bm{x},t)],
\end{align}
where $\Delta t$ is the time step, and $\tau_f = \displaystyle\frac{\tau}{\Delta t}$. The time step is the same as the lattice unit.
Similarly, we obtain the discretized evolution equation of $h$,
\begin{align}\label{Eq:DiscreteReductionH}
h_i(\bm{x}+ \bm{\xi}_i &\Delta t,t + \Delta t)\notag\\
=&h_i(\bm{x},t) -  \frac{Pr}{\tau_f}[h_i(\bm{x},t) -\frac{N_f}{2}T G_i(\bm{x},t)].
\end{align}
From Eqs.~(\ref{Eq:DiscreteReductionG}) and (~\ref{Eq:DiscreteReductionH}), we can derive the Navier-Stokes equations, with which the Prandtl number is defined by Eq.~(\ref{Eq:Pr}) and the specific heat ratio is defined by Eq.~(\ref{Eq:Gamma}).
Eq.~(\ref{Eq:DiscreteReductionG}) and ~(\ref{Eq:DiscreteReductionH}) are employed to update the discrete distribution functions, i.e. $g_i$ and $h_i$.

\section{Numerical validation}
In this section, the thermal Couette flow and the one-dimensional shock tube flow are carried out to verify the LB scheme proposed in this work.

\subsection{Thermal Couette flow}\label{Couette}
The analytical temperature distribution along the $y$-direction of the thermal Couette flow in a steady state is given by
\begin{equation}\label{Eq:Couette}
T=T_0 + \frac{Pr}{2C_p}u_0^2\frac{y}{L_0}(1-\frac{y}{L_0}),
\end{equation}
where $L_0$ is the distance between the upper plate and the lower plate, $y$ is the distance from a point to the lower plate, $u_0$ is the $x$-direction velocity of the top plate at the beginning. The dimensionless variables are defined by
\[ \begin{array}{ccc}
\tilde y = \displaystyle\frac{y}{L_0},
\tilde u_0 = \displaystyle\frac{u_0}{\sqrt{\theta_0}},
\tilde T = \displaystyle\frac{T}{T_0},
\end{array}\]
where $\theta_0\!=\!R_gT_0$. Inserting the dimensionless variables $\tilde y$, $\tilde u$, $\tilde T$ and the specific heat on constant pressure $C_p = \displaystyle \frac{D+N_f+2}{2}R_g$ into Eq(~\ref{Eq:Couette}), omitting the tildes, we obtain the dimensionless form of Eq(~\ref{Eq:Couette}):
\begin{equation}\label{Eq:CouetteDi3}
 T = 1 + u_0^2\frac{Pr}{D+N_f+2}  y (1 -  y).
\end{equation}
The specific heat ratio is defined by $\gamma\!=\!\displaystyle\frac{D+N_f+2}{D+N_f}$, so $\gamma$ can be modified by $N_f$.

The analytical velocity distribution along the $y$-direction of the thermal Couette flow is given by
\begin{align}\label{Eq:CouetteVelocity}
u(t) &= \frac{y}{L_0}u_0 \notag\\
&+ \frac{2}{\pi}u_0\sum_{n=1}^{\infty}\left[\frac{(-1)^n}{n}\exp\left(-n^2\pi^2\frac{\mu t}{\rho L_0^2}\right)\sin\left(\frac{n\pi y}{L_0}\right)\right].
\end{align}
The dimensionless variables are defined by
\[ \begin{array}{ccc}
\tilde \mu = \displaystyle\frac{\mu}{\rho_0 L_0 \sqrt{(RT_0)}},
\tilde t = \displaystyle\frac{t\sqrt{(RT_0)}}{L_0},
\tilde \rho = \displaystyle\frac{\rho}{\rho_0}.
\end{array}\]
Inserting $\tilde \mu$, $\tilde t$, $\tilde \rho$, and other dimensionless variables defined hereinabove into Eq.~(\ref{Eq:CouetteVelocity}), substituting the definition of $\mu = \tau p Pr$, and omitting the tildes, we obtain the dimensionless form of Eq.~(\ref{Eq:CouetteVelocity}):
\begin{equation}\label{Eq:CouetteVelocityDimensionless}
u(t) = y + \frac{2}{\pi}\sum_{n=1}^{\infty}\left[\frac{(-1)^n}{n}\exp\left(-n^2 \pi^2 \tau Pr T t\right)\sin\left(n \pi y\right)\right].
\end{equation}

We set the initial conditions as $\rho=1$, $T=1$, $u_x=u_y=0$. Here, $u_x$ is the $x$-directional velocity and $u_y$ is the $y$-directional velocity. The top plate moves initially with the velocity $u_0=1$. The grid is $X \times Y=50 \times 400$. We also set $\nu=0.5$, $\kappa=0.2$, and $N_f=3$, so the Prandtl number is $Pr=0.71$ and the specific heat ratio is $\gamma=1.4$. A lattice model, named D2Q37, is employed. We design the lattice model employing the Hermite quadrature \cite{philippi2006continuous,philippi2015high,mattila2014high,shan2006kinetic,shan2010general,shim2013multidimensional,shim2013univariate}.  D2Q37 is of fourth-order accuracy. The discrete particle velocity set and the weights $\omega_i$ of D2Q37 are shown in Table~~\ref{tab:D2Q37}.
 \begin{table}
\caption{  Discrete velocities and weights of D2Q37.
$Perm$ denotes permutation and $k$ denotes the number of
 discrete velocities included in each group. The scaling factor is $r =\! 1.19697977$.}
 \begin{ruledtabular}
\begin{tabular}{ p{30pt} p{60pt}  p{80pt}   }

 $k$        & $\bm{\xi}_i$   & $\omega_i$      \\
\hline
  $1 $ & $(0,0)$       & $2.03916918e\!-\!1$      \\
  $4 $ & $Perm(r,0)$   & $1.27544846e\!-\!1$      \\
  $4 $ & $Perm(r,r)$   & $4.37537182e\!-\!2$      \\
  $4 $ & $Perm(2r,0)$  & $8.13659044e\!-\!3$      \\
  $8 $ & $Perm(2r,r)$  & $9.40079914e\!-\!3$      \\
  $4 $ & $Perm(3r,0)$  & $6.95051049e\!-\!4$      \\
  $8 $ & $Perm(3r,r)$  & $3.04298494e\!-\!5$      \\
  $4 $ & $Perm(3r,3r)$ & $2.81093762e\!-\!5$      \\
\end{tabular}
\end{ruledtabular}
\label{tab:D2Q37}
\end{table}
The periodic boundary condition is applied to the left and right sides, while a hybrid boundary condition is applied to the up and down boundaries. This hybrid scheme consists of two parts: the equilibrium part, which is obtained through the kinetic boundary condition (KBC)\cite{ansumali2002kinetic,sofonea2009implementation}, and the non-equilibrium part, which is the same as that of the non-equilibrium extrapolation scheme (NEEP)\cite{zhao2002non}. Notably, the density of the wall nodes is obtained through KBC. One of the advantages of the hybrid boundary scheme is that it eliminates velocity slip and temperature slip. We will discuss the hybrid boundary scheme in more detail in future work.
Fig.~\ref{Fig:PrAndGamma} shows the temperature distribution along the $y$ coordinate in a steady state. Fig.~\ref{Fig:PrAndGamma:Vel} shows the velocity distribution along the $y$ coordinate at times $t\!=\!5$, $t\!=\!10$, $t\!=\!20$, and $t\!=\!60$. The numerical solutions are in  agreement with the analytical solutions.
\begin{figure}
\includegraphics[width=7cm]{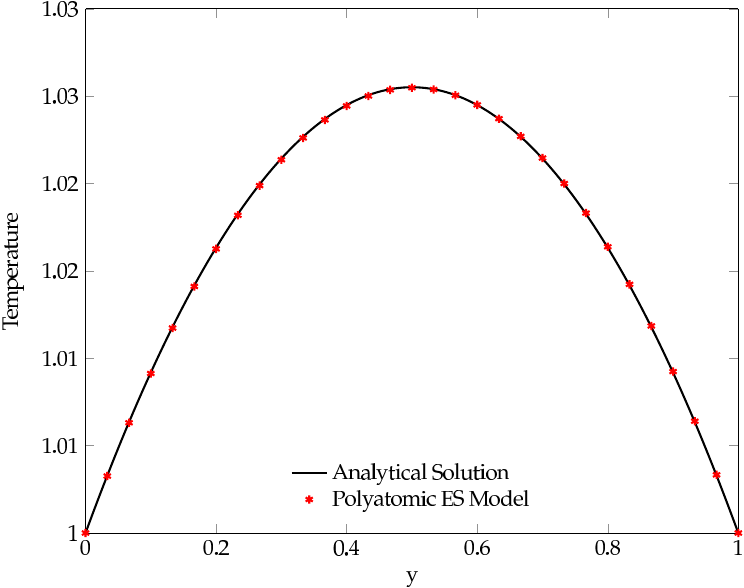}
\caption{\small{Thermal Couette flow.The temperature distribution along $y$ coordinates in steady state.}
}
\label{Fig:PrAndGamma}
\end{figure}
\begin{figure}
\includegraphics[width=7cm]{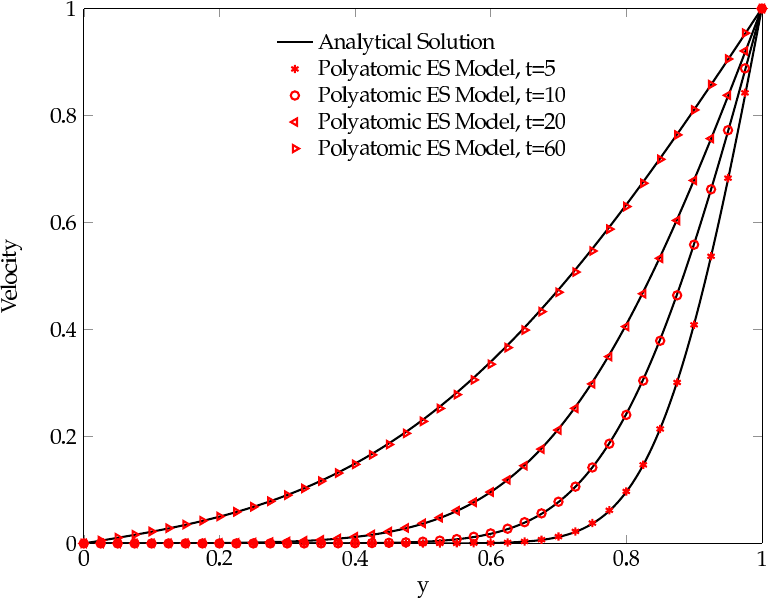}
\caption{\small{Thermal Couette flow. The velocity distribution along $y$ coordinate at time $t=5$, $t=10$, $t=20$ and $t=60$.
}
}
\label{Fig:PrAndGamma:Vel}
\end{figure}

\subsection{The Shock tube flow of 1D}
The shock tube problem of one dimension has been discussed extensively in the literature \cite{SOD19781}. The initial conditions are given by $(\rho_L,T_l, u_{lx}) = (4,1,0)$ and $(\rho_R,T_r, u_{rx}) = (1,1,0)$ on the left and right sides of the shock tube, respectively, where $u_{lx}$ and $u_{rx}$ are the macroscopic velocities along the $x$ coordinate. The specific heat ratio is set as $\gamma=1.4$ and the relaxation time as $\tau=2/3$. The grid is $X \times Y = 1000 \times 16 $, with additional degrees of freedom $N_f=3$. The periodic boundary condition is employed for the upper and lower boundaries, while the open boundary condition is used for the left and right boundaries.

The D2Q33 lattice model proposed by J. Shim\cite{shim2013multidimensional} is employed. Of all the existing lattice models of fourth-order accuracy, this lattice model has the least discrete velocities, making it more efficient than others. Table~\ref{tab:D2Q33} gives the discrete velocity set and the weight coefficients of the D2Q33 model.
\begin{table}
\caption{  Discrete velocities and weights of D2Q33.
$Perm$ denotes permutation and $k$ denotes the number of
 discrete velocities included in each group. The scaling factor is $r \!=\! 0.819381\times  \sqrt{2}$.}
 \begin{ruledtabular}
\begin{tabular}{ p{30pt} p{60pt}  p{80pt}   }
 $k$        & $\bm{\xi}_i$   & $\omega_i$      \\
\hline
  $1 $ & $(0,0)$       & $1.61987e\!-\!1$      \\
  $4 $ & $Perm(r,0)$   & $1.43204e\!-\!1$      \\
  $4 $ & $Perm(r,r)$   & $3.38840e\!-\!2$      \\
  $4 $ & $Perm(2r,0)$  & $5.56112e\!-\!3$      \\
  $4 $ & $Perm(2r,2r)$  & $8.44799e\!-\!5$     \\
  $4 $ & $Perm(3r,0)$  & $1.13254e\!-\!3$      \\
  $8 $ & $Perm(2r,r)$  & $1.28169e\!-\!2$      \\
  $4 $ & $Perm(4r,4r)$ & $3.45552e\!-\!6$      \\
\end{tabular}
\end{ruledtabular}
\label{tab:D2Q33}
\end{table}

Fig.~\ref{Fig:HeatRatio2D} presents the simulation results at $step$ 174, which corresponds to a time of $t = 0.15$ when expressed in terms of $\Delta t$, $X$, and $r$. This can be calculated as follows: $t = \frac{step\times \Delta t}{ X \times r} = \frac{174\times 1}{1000\times 0.819381 \sqrt{2} } \approx 0.15$.
\begin{figure*}
\includegraphics[width=14cm]{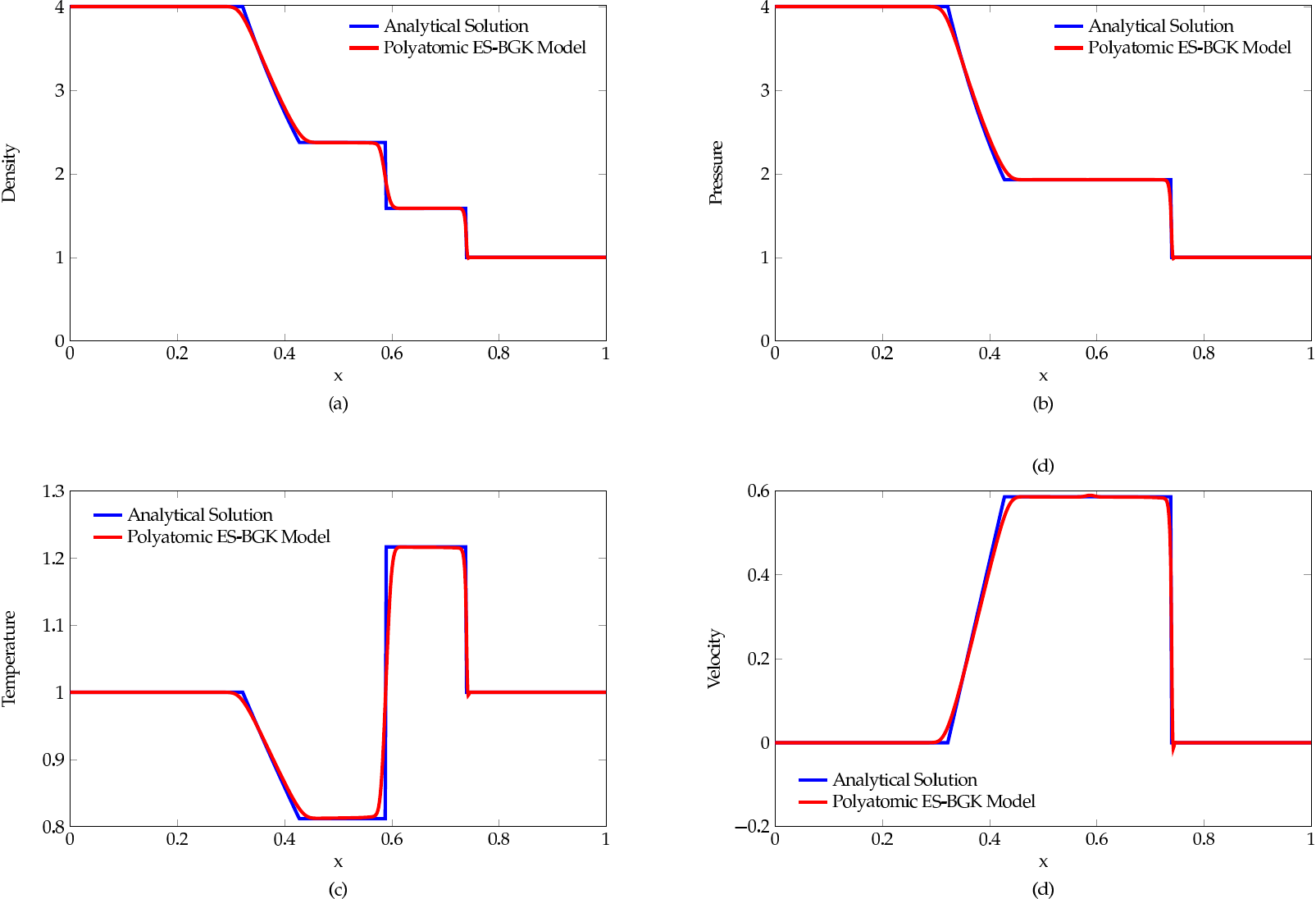}
\caption{\small{The simulation results of D2Q33 and the analytical resolutions  at the time $t\!\approx\!0.15$. The specific heat ratio is $\gamma\! = \! 1.4$ and the parameter $N_f$ is $3$. The relaxation time is $\tau\!=\! 2/3$. The initial condition of the left side is ${\rho \!=\! 4,T\!=\!1,\bm{u}\!=\!0}$ and that of the right side is ${\rho \!=\! 1,T\!=\!1,\bm{u}\!=\!0}$.}}
\label{Fig:HeatRatio2D}
\end{figure*}
The results are in agreement with the analytical resolutions.

\subsection{A two dimensional supersonic flow on a blunt flat plate}

Simulating two-dimensional supersonic flow on a flat blunt plate with an inlet velocity of $u_{inlet} = 3 \text{Ma}$, the NND\cite{zhang1991nnd} scheme is used to discretize the advection term of Eq.~(\ref{Eq:DisReductionHA}), while the Euler method is employed to discretize the temporal term of Eq.~(\ref{Eq:DisReductionHA}). 

The velocity boundary condition is applied to the inlet boundary, and the hybrid boundary condition discussed previously is applied to all the walls of the flat plate. The wall density is computed in the same way as the kinetic boundary condition (KBC).The outlet boundary, the upper boundary, and the lower boundary of the entire computational area are set to the zero-gradient boundary condition. 

The initial conditions are set as follows: the internal velocity of $\bm{u} = 0$, density of $\rho = 1$, and temperature of $T = 0.71429$, all of which are dimensionless. A computational grid of $200 \times 400$ cells is employed, with 60 cells between the front wall of the blunt flat plate and the inlet, 80 cells between the upper wall of the blunt flat plate and the upper boundary of the entire computational area, and the same number of cells between the lower wall of the blunt flat plate and the lower boundary. 

The Courant–Friedrichs–Lewy (CFL) principle is employed to determine the time step, with the Courant
number set to 0.8. The collision frequency of gas molecules is set to $\gamma = 1/6$, and the D2Q37   model discussed previously  is utilized.

The pressure configurations at different times are depicted in Fig.~(\ref{Fig:BluntFlow}). In the final figure, the flow becomes stable.

\begin{figure}[htbp]
	\centering
	\subfloat{\includegraphics[width=4cm]{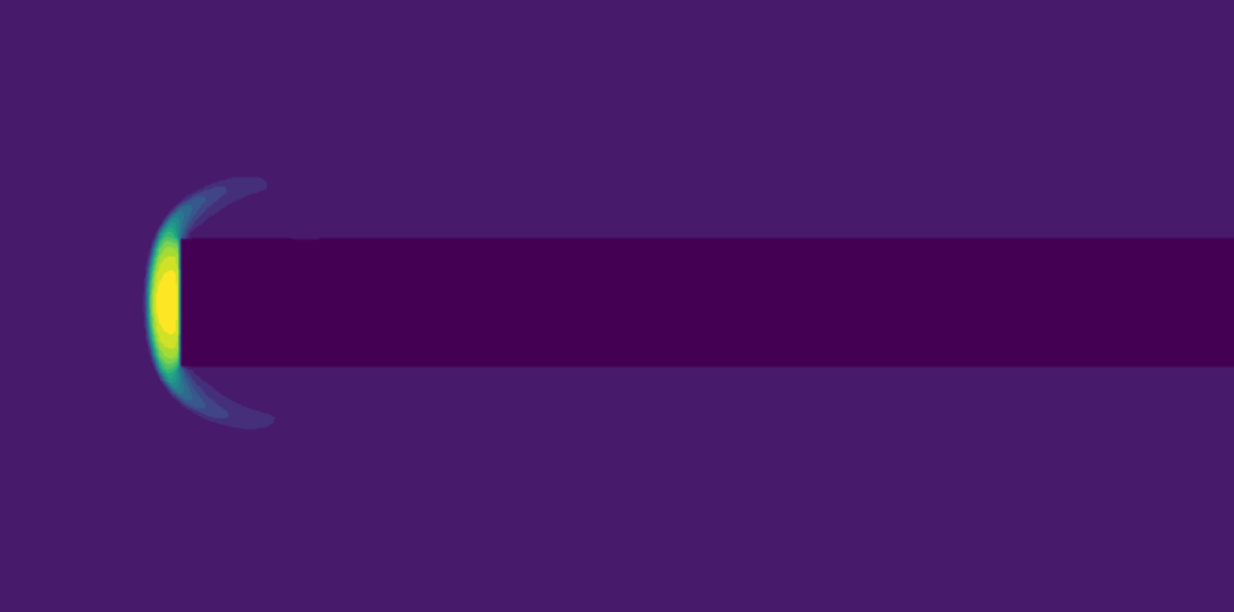}}\quad
	\subfloat{\includegraphics[width=4cm]{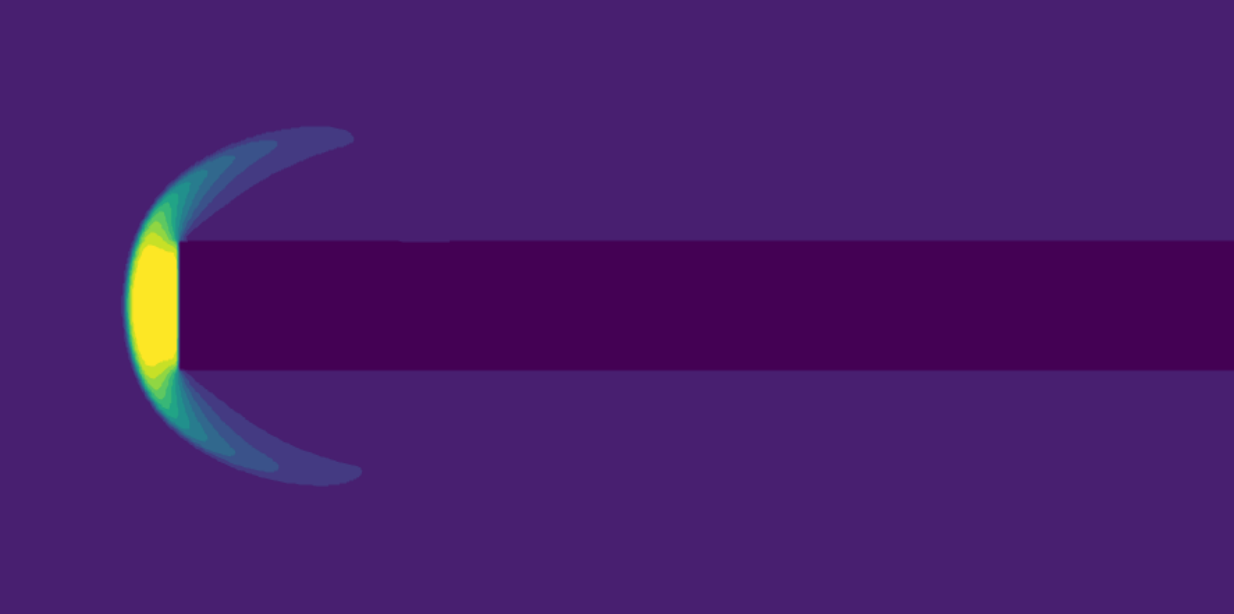}}\\
	\subfloat{\includegraphics[width=4cm]{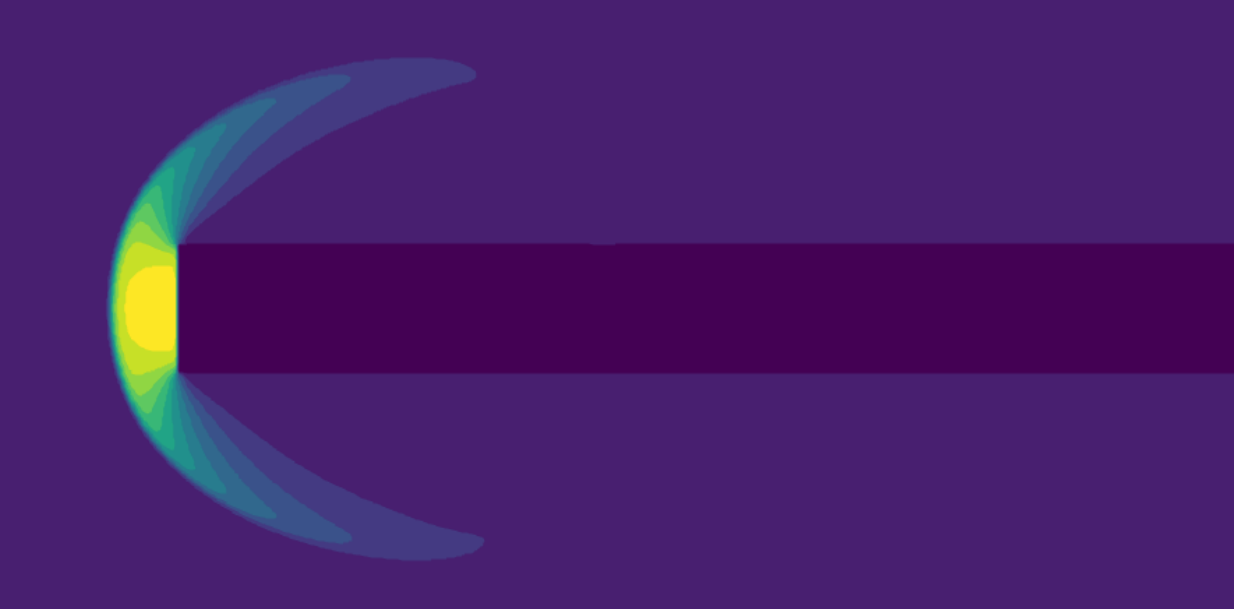}}\quad
	\subfloat{\includegraphics[width=4cm]{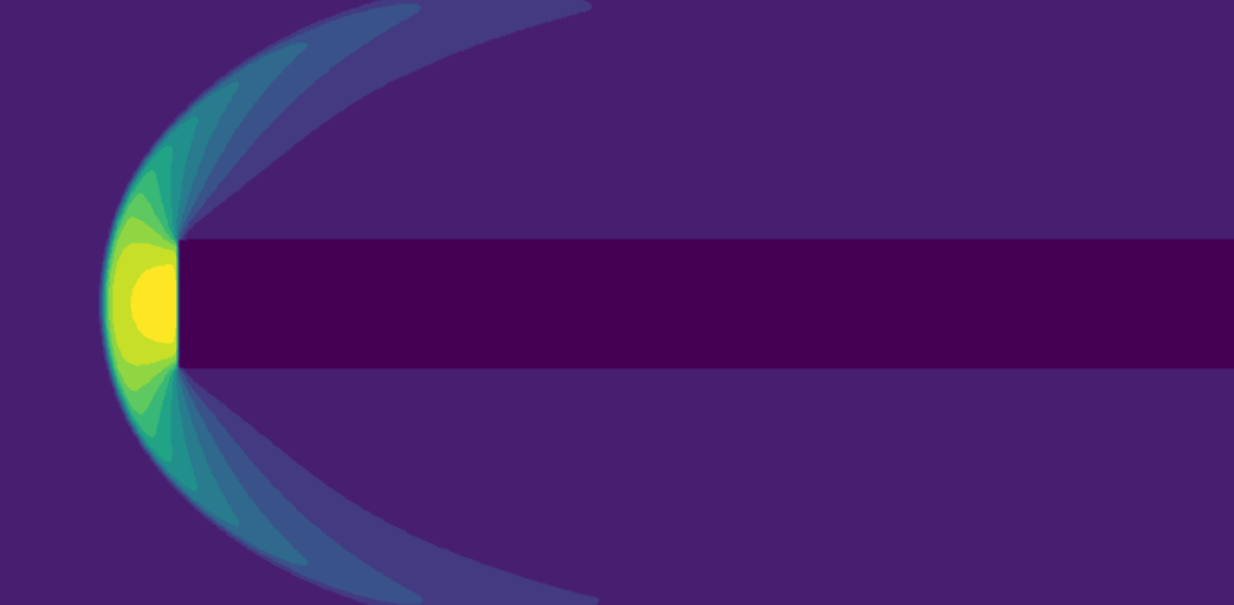}}\\
	\subfloat{\includegraphics[width=4cm]{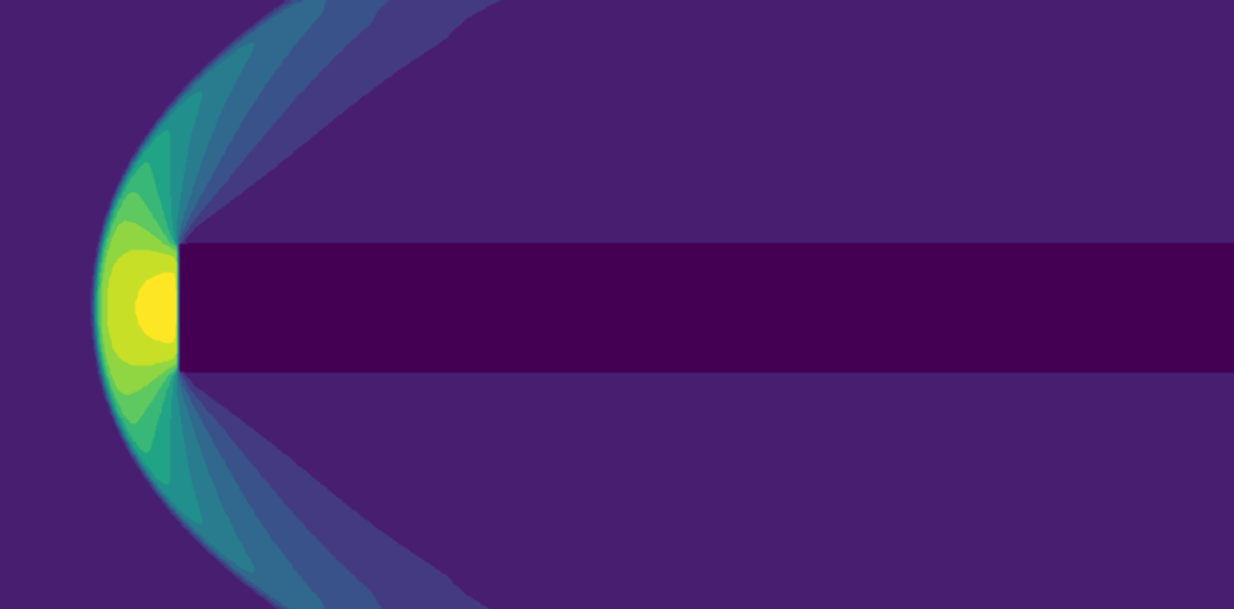}}\quad
	\subfloat{\includegraphics[width=4cm]{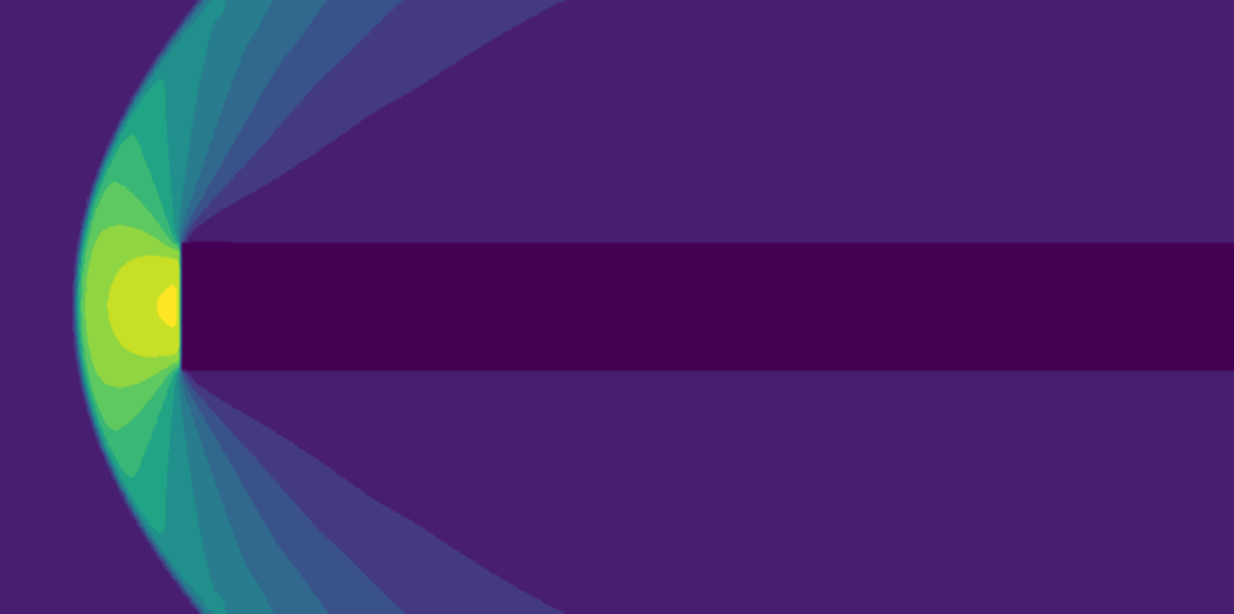}}
	\caption{\small{The pressure configuration of the two-dimensional supersonic flow on a blunt flat plate at various points in time}}
	\label{Fig:BluntFlow}
\end{figure}

\section{Conclusion}
A lattice Boltzmann scheme based on the polyatomic ES-BGK model is proposed, from which the Navier-Stokes equations  with an arbitrary  Prandtl number and specific heat ratio can be derived via the Chapman-Enskog expansion. 
The Gaussian distribution is expanded on the Hermite polynomials, and the general term formula for the Hermite coefficients of the Gaussian distribution is deduced, which is the main attribution of this work. To verify the scheme, the thermal Couette flow and  the shock tube flow of one dimension are simulated, and the results agree well with the analytical resolutions. A two dimensional supersonic flow on a blunt flat palte is also simulated.
The proposed scheme offers a novel way to modify the Prandtl number and the specific heat ratio for the lattice Boltzmann method. As the ES-BGK model satisfies the entropy principle, the lattice Boltzmann scheme proposed in this work can prevent the non-physical results caused by the violation of the $H$-theorem.

\appendix
\section{The derivation of Eq.~(\ref{Eq:fESCoe})}\label{appendix:A}
The derivation of Eq.~(\ref{Eq:fESCoe}) is as follows:
\begin{align}
 \bm{a}^{(n)} 
 =& \frac{1}{\rho}\int \bm{H}^{(n)}(\bm{u} + \bm{\eta}\cdot \bm{\Lambda}^{\frac{1}{2}})\frac{\rho}{\sqrt{det(2\pi \bold{\Lambda})}}\notag\\
 &\times\exp [-\frac{1}{2}(\bm{\xi} - \bm{u})\cdot \bm{\Lambda}^{-1} \cdot (\bm{\xi} -\bm{u})] d (\bm{u} + \bm{\eta} \cdot \bm{\Lambda}^{\frac{1}{2}} )\notag\\
 = & \int \bm{H}^{(n)}(\bm{u} + \bm{\eta}\cdot \bm{\Lambda}^{\frac{1}{2}})\omega(\bm{\eta}) d \bm{\eta} \notag\\
 = &\sum_{\alpha=0}^{[n/2]}(-1)^{\alpha}\int \omega(\bm{\eta}) (\bm{u} + \bm{\eta} \cdot \bm{\Lambda}^{\frac{1}{2}})^{n-2\alpha} \bm{\delta}^{\alpha} d \bm{\eta} \notag \\
  = &\sum_{\alpha=0}^{[n/2]} \sum_{\beta =0}^{[n/2]-\alpha}  (-1)^{\alpha} \int \omega(\bm{\eta}) \bm{u}^{n-2(\alpha +\beta)} (\bm{\eta} \cdot \bm{\Lambda}^{\frac{1}{2}})^{2\beta} \bm{\delta}^{\alpha} d \bm{\eta} \notag \\
= &\sum_{\alpha=0}^{[n/2]} \sum_{\beta =0}^{[n/2]-\alpha}  (-1)^{\alpha} \bm{u}^{n-2(\alpha +\beta)} \bm{\delta}^{\alpha}
   [\int \omega(\bm{\eta}) \bm{\eta}^{2\beta} d \bm{\eta} \cdot \bm{\Lambda}^{\beta}]  \notag  \\ 
   = &\sum_{\alpha=0}^{[n/2]} \sum_{\beta =0}^{[n/2]-\alpha}  (-1)^{\alpha}
  \bm{\delta}^{\alpha} (\bm{\delta}^{\beta}\cdot \bm{\Lambda}^{\beta} )\bm{u}^{n-2(\alpha +\beta)} \notag\\
= &\sum_{\alpha=0}^{[n/2]} \sum_{\beta =0}^{[n/2]-\alpha}  (-1)^{\alpha}  \bm{\delta}^{\alpha} \bm{\Lambda}^{\beta} \bm{u}^{n-2(\alpha +\beta)}.\notag
\end{align}

\section{The expansion of Eq.~(\ref{Eq:GExpansion})}\label{appendix:B}
The expansion of Eq.~(\ref{Eq:GExpansion}) is as follows:
\begin{eqnarray}
\bm{a}^{(0)}\cdot \bm{H}^{(0)}  =&& 1,\notag\\
\bm{a}^{(1)}\cdot \bm{H}^{(1)}  = &&u_i \xi_i,\notag\\
\bm{a}^{(2)}\cdot \bm{H}^{(2)}  =&& \Lambda_{ij}\xi_i\xi_j-\Lambda_{ii}-\xi^2+D+(u_i\xi_i)^2-u^2, \notag\\
\bm{a}^{(3)}\cdot \bm{H}^{(3)}  =&& 3\Lambda_{ij}\xi_i\xi_j(u_i \xi_i)\notag\\
&&+ 3[u^2(\Lambda_{ij}\xi_i\xi_j)+2(u_i u_i)(\Lambda_{jk}u_j\xi_k)]\notag\\
&&+ (u_i \xi_i)^3 + 3(u_i \xi_i)^2\xi^2 - 3(u_i\xi_i)^2u^2\notag\\
&&-3(D+2)(u_i \xi_i),\notag\\
\bm{a}^{(4)}\cdot \bm{H}^{(4)}
=&& 3\Lambda_{ij}\xi_i\xi_j - 6(\Lambda_{ij}\xi_i\xi_j \notag\\
&&+2\Lambda_{ik}\xi_i\lambda_{jk}\xi_j)\notag\\
&&+6(\Lambda_{ii}^2+2\Lambda_{ij}\Lambda_{ij})\notag\\
&&+6\Lambda_{ij}\xi_i\xi_j\xi^2-6[(D+4)\Lambda_{ij}\xi_i\xi_j+\Lambda_{ii}\xi^2]\notag\\
&&+6(D+2)\Lambda_{ii}\notag+6\Lambda_{ij}\xi_i\xi_j(u_k\xi_k)^2\notag\\
&&-6[\Lambda_{ij}\xi_i\xi_j u^2 \notag\\
&&+ 4\Lambda_{ij}\xi_i u_j (u_k \xi_k)+\Lambda_{ii} (u_k \xi_k)^2 ]\notag\\
&&+6(\Lambda_{ii}u^2+ \Lambda_{ij}u_i u_j )\notag\\
&&+ (\xi_i u_i)^4 - 6(\xi_i u_i)^2 u^2 + 3u^4  \notag\\
&&- 6[(\xi_i u_i)^2(u^2 - D - 4)
 + (D + 2 - u^2 )\xi^2] \notag\\
 &&+ 3[u^4 - 2(D+2)u^2+D(D+2)]. \notag
\end{eqnarray}

\nocite{*}

\bibliography{RaPr}

\end{document}